\documentclass[12pt]{article}
\pdfoutput=1
\usepackage{graphicx}
\usepackage{caption}
\usepackage{subcaption}
\usepackage{epsfig}
\usepackage{rotating}
\usepackage{amsmath}
\usepackage{feynmp}
\def\br(#1,#2){\left\langle#1#2\right\rangle}
\def\sq(#1,#2){\left[#1#2\right]}
\def\s(#1,#2){s_{#1 #2}}
\def\t(#1,#2,#3){s_{#1 #2 #3}}

\begin{document}

\begin{titlepage}
\begin{flushright}
	{\today} \\
	MPP-2013-101 \\
	CP3-13-13
\end{flushright}
 \vskip2cm
\begin{center}
{\Large \bf Effective Field Theory of \\\vspace{12pt} Precision Electroweak Physics at One Loop} \\
\medskip
\bigskip\bigskip\bigskip\bigskip
{\bf Harrison Mebane$^1$, Nicolas Greiner$^{1,2}$, Cen Zhang$^{1,3}$, and Scott Willenbrock$^1$ }\\
\bigskip\bigskip\bigskip
$^1$Department of Physics, University of Illinois at Urbana-Champaign \\ 1110 West Green Street, Urbana, IL  61801 \\
\bigskip
$^2$Max-Planck-Institut f\"ur Physik, F\"ohringer Ring 6, 80805 M\"unchen, Germany \\
\bigskip
$^3$Centre for Cosmology, Particle Physics and Phenomenology (CP3) \\ Universit\'e Catholique de Louvain, B-1348
Louvain-la-Neuve, Belgium
\end{center}

\bigskip\bigskip\bigskip

\begin{abstract}
The one loop effects of two dimension-six operators on gauge boson self energies are computed within an effective field theory framework.  These self energies are translated into effects on precision electroweak observables, and bounds are obtained on the operator coefficients.  The effective field theory framework allows for the divergences that arise in the loop calculations to be properly handled, and for unambiguous bounds on the coefficients to be obtained.  We find that the coefficients are only weakly bounded, in contrast to previous calculations that obtained much stronger bounds.  We argue that the results of these previous calculations are specious.
\end{abstract}

\end{titlepage}

In addition to searching for direct evidence of new physics, data can be probed for the indirect effects of new heavy particles.  The most general, model-independent framework for considering the indirect effects of new physics is effective field theory \cite{Weinberg:1978kz}.  The new physics is parameterized by effective operators not present in the Standard Model. The power of effective field theory is particularly on display when dealing with one-loop calculations.  The effective field theory framework provides a systematic means to deal with the divergences that arise in loop calculations, yielding a finite and unambiguous result.

In this paper, we continue an analysis begun in Refs.~\cite{De Rujula:1991se,Hagiwara:1992eh,Hagiwara:1993ck}
on the loop-level effects of effective operators on precision electroweak observables. Those papers focused on the divergent portions of the loop diagrams and a subset of the finite parts, and did not appreciate that unambiguous bounds could be obtained on the coefficients of the effective operators.  We use the full (finite plus divergent) expressions in order to obtain unambiguous bounds on a particular pair of effective operator coefficients.  These calculations involve only weak boson self energies, also called oblique corrections.  Methods for organizing and applying such corrections to observables are well known \cite{Kennedy:1988sn,Peskin:1991sw} and will be used throughout the analysis.

Because we are dealing with loops of gauge or Higgs bosons, which are not significantly heavier than the $W$ mass, the $S$, $T$, $U$, parametrization of Ref.~\cite{Peskin:1991sw} may not be accurate.  This is in contrast to an analysis of top-quark loops in which the $S$, $T$, $U$, parameters are useful due to the large mass of the top quark \cite{Greiner:2011tt}.  Here we instead use the ``star'' formalism of Ref.~\cite{Kennedy:1988sn}, as in Ref.~\cite{Zhang:2012cd} which involves both top and bottom loops.

When used to extend the Standard Model, a general effective field theory can be written in the form
\begin{equation}
	\mathcal{L}_{eff} = \mathcal{L}_{SM} + \sum_i \frac{c_i}{\Lambda^2}\mathcal{O}_i + \ldots
\end{equation}
where the $c_i$ are dimensionless coefficients, $\Lambda$ is the energy scale of new physics, and the $\mathcal{O}_i$ are effective operators with mass dimension six.  We have not included a dimension-five term in the above expression because there is only one such operator, and it does not involve weak bosons \cite{Weinberg:1979sa}.

In the basis of Ref.~\cite{Hagiwara:1993ck}, there are nine dimension-six operators involving only weak bosons and/or Higgs doublets that affect precision electroweak measurements at tree or one-loop level.  Four of these operators affect weak boson self energies at tree level.  Of the remaining five, three affect triple gauge couplings and can be bounded at tree level from weak boson pair production at high-energy colliders.  This paper will focus on the remaining two operators,
\begin{equation}
\begin{aligned}
	\mathcal{O}_{WW} & = \phi^\dagger \hat{W}^{\mu\nu}\hat{W}_{\mu\nu} \phi \\
	\mathcal{O}_{BB} & = \phi^\dagger \hat{B}^{\mu\nu}\hat{B}_{\mu\nu} \phi
\end{aligned}
\label{loopops}
\end{equation}
where $\hat{B}_{\mu\nu} = \frac{ig^\prime}{2}B_{\mu\nu}$, $\hat{W}_{\mu\nu} = ig\frac{\sigma^{a}}{2} W^{a}_{\mu\nu}$ ($\sigma^a$ are the Pauli matrices), and $\phi$ is the Standard Model Higgs doublet.

The above two operators affect precision electroweak observables only through oblique corrections.  When the Higgs field takes its vacuum expectation value, both operators appear to affect the weak boson self energies at tree level; however, because the operators have the same form as the Standard Model gauge kinetic terms, all corrections generated by these operators that involve only gauge bosons can be absorbed into the Standard Model through field and coupling redefinitions.  Thus, the only observable corrections from these operators arise from effective interactions involving gauge bosons and at least one Higgs boson or Goldstone boson.  For this reason, these operators cannot be bounded from processes involving only vector bosons.

Explicitly, the new interactions generated by the operators $\mathcal{O}_{WW}$ and $\mathcal{O}_{BB}$ that contribute at the one-loop level are
\begin{align}
	\mathcal{L}_{eff} & = \mathcal{L}_{SM} - \frac{c_{WW}}{\Lambda^2} \frac{g^2}{8}\left(2W^+_{\mu\nu} W^{-\mu\nu}
	 + c^2 Z_{\mu\nu} Z^{\mu\nu} + s^2 A_{\mu\nu} A^{\mu\nu} + 2s c A_{\mu\nu} Z^{\mu\nu}\right) \\
	& \hspace{2in} \times \left(2\phi^+ \phi^- + 2v H + H H +
	  \phi^0 \phi^0\right) \nonumber \\
	& \qquad \qquad - \frac{c_{BB}}{\Lambda^2} \frac{g^{\prime 2}}{8}\left(s^2 Z_{\mu\nu} Z^{\mu\nu}
	+ c^2 A_{\mu\nu} A^{\mu\nu} - 2s c A_{\mu\nu} Z^{\mu\nu}\right) \nonumber \\
	& \hspace{2in} \times \left(2\phi^+ \phi^- + 2v H + H H +
	  \phi^0 \phi^0\right) \nonumber
\end{align}
These interactions induce several contributions to the gauge boson self energies at the one-loop level.  The general structure of the relevant diagrams appears in Figure \ref{fig:loops}, and the explicit self energies can be found in Appendix \ref{apdx:corrections}.
\begin{figure}
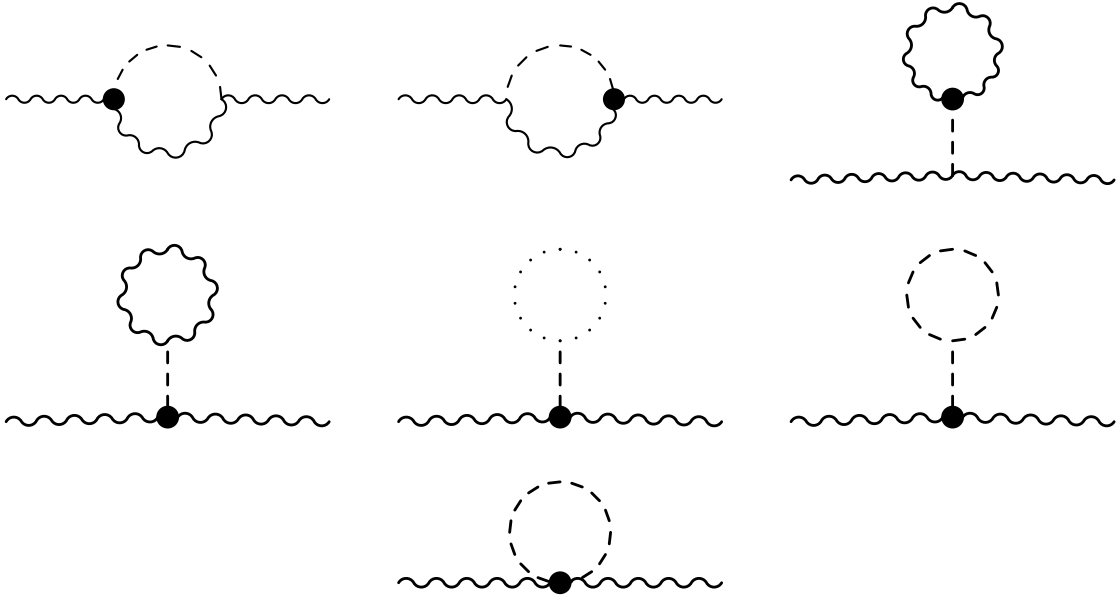

\centering
\begin{minipage}[t]{2in}
\begin{center}
\includegraphics[width=1.7in]{loop1} \\
\vspace{14pt}
\includegraphics[width=1.7in]{loop4}
\end{center}
\end{minipage}
\begin{minipage}[t]{2in}
\begin{center}
\includegraphics[width=1.7in]{loop2} \\
\vspace{14pt}
\includegraphics[width=1.7in]{loop5} \\
\vspace{14pt}
\includegraphics[width=1.7in]{loop7}
\end{center}
\end{minipage}
\begin{minipage}[t]{2in}
\begin{center}
\includegraphics[width=1.7in]{loop6} \\
\vspace{14pt}
\includegraphics[width=1.7in]{loop3}
\end{center}
\end{minipage}
\caption{Loop-level contributions of the operators $\mathcal{O}_{WW}$ and $\mathcal{O}_{BB}$.  Wavy lines represent gauge bosons, dashed lines represent Higgs or Goldstone bosons, dotted lines represent ghost fields, and the black dots represent effective operator interactions.}
\label{fig:loops}
\end{figure}

The self energies contain divergences that must be eliminated in order to arrive at meaningful results.  Because of the gauge-invariant structure of the effective field theory, divergences arising from operators of a given dimension can always be absorbed by some other operator of the same dimension.  As shown in Ref.~\cite{Hagiwara:1993ck}, the operator
\begin{equation}
	\mathcal{O}_{BW} = \phi^\dagger \hat{B}^{\mu\nu} \hat{W}_{\mu\nu} \phi
\end{equation}
contributes to gauge boson self energies at tree level and is able to absorb all oblique divergences arising from the operators $\mathcal{O}_{WW}$ and $\mathcal{O}_{BB}$.  Thus the operator $\mathcal{O}_{BW}$ must be included in our analysis.  An analogous situation arises for top-quark loops \cite{Greiner:2011tt}.

Electroweak boson self energies contribute to precision electroweak data through corrections to the input variables $\alpha$, $m_Z$, and $s^2$.  The correction to $\alpha$ depends upon the type of vertex; these corrections will be labeled $\delta\alpha_\gamma$, $\delta\alpha_Z$, or $\delta\alpha_W$, depending on the mediating boson.  The self energy between bosons $X$ and $Y$ is denoted $\Pi_{XY}$ in the expressions below:
\begin{align}
	\alpha + \delta\alpha_\gamma & = \alpha\left(1+\Pi_{\gamma\gamma}^\prime(q^2)
		-\Pi_{\gamma\gamma}^\prime(0)\right) \\
	\alpha + \delta\alpha_Z & = \alpha\left(1+\Pi_{\gamma\gamma}^\prime(q^2)
		-\Pi_{\gamma\gamma}^\prime(0)\right) \\
	& \qquad \qquad \times \left(1+\frac{d}{dq^2}\Pi_{ZZ}(m_Z^2)-\Pi_{\gamma\gamma}^\prime(q^2)
		-\frac{c^2-s^2}{c s}\Pi_{\gamma Z}^\prime(q^2)\right) \nonumber \\
	\alpha + \delta\alpha_W & = \alpha\left(1+\Pi_{\gamma\gamma}^\prime(q^2)
		-\Pi_{\gamma\gamma}^\prime(0)\right) \\
	& \qquad \qquad \times \left(1+\frac{d}{dq^2}\Pi_{WW}(m_W^2)-\Pi_{\gamma\gamma}^\prime(q^2)
		-\frac{c}{s}\Pi_{\gamma Z}^\prime(q^2)\right) \nonumber \\
	m_Z^2 + \delta m_Z^2 & = m_Z^2 - \Pi_{ZZ}(m_Z^2) + \Pi_{ZZ}(q^2)
		- (q^2-m_Z^2)\frac{d}{dq^2}\Pi_{ZZ}(m_Z^2) \\
	s^2 + \delta s^2 & = s^2\left[1-\frac{c}{s}\Pi_{\gamma Z}^\prime - \frac{c^2}{c^2-s^2}
		\left(\Pi_{\gamma\gamma}^\prime(0) + \frac{1}{m_W^2}\Pi_{WW}(0)-\frac{1}{m_Z^2}\Pi_{ZZ}(m_Z^2)\right)\right]
\end{align}
where $\Pi_{XY}^\prime(q^2) = (\Pi_{XY}(q^2)-\Pi_{XY}(0))/q^2$.  Explicit expressions for the self energies are given in Appendix~\ref{apdx:corrections}.

The correction to any electroweak observable $X$ measured at an energy at or above the $Z$-pole is given by
\begin{equation}
	\delta X = \frac{\delta X}{\delta \alpha} \delta\alpha + \frac{\delta X}{\delta m_Z^2} \delta m_Z^2 +
		\frac{\delta X}{\delta s^2} \delta s^2
	\label{dx1}
\end{equation}
In contrast, low-energy observables are affected by corrections to $s^2$ and by changes to the $\rho$ parameter
\begin{equation}
	\delta X = \frac{\delta X}{\delta s^2} \delta s^2 + \frac{\delta X}{\delta \rho} \delta \rho
	\;.
	\label{dx2}
\end{equation}
where $\delta \rho = \frac{1}{m_W^2}\Pi_{WW}(0) -	\frac{1}{m_Z^2}\Pi_{ZZ}(0)$ and the Standard Model value of $\rho$ is unity.  

In our calculations, we used the following values for input parameters:
\begin{align}
	\alpha(m_Z)  = 1/128.91, \qquad v & = 246.2\textrm{ GeV}, \qquad m_Z = 91.1876\textrm{ GeV}, \qquad
	m_h = 125\textrm{ GeV} \\
	m_t = 172.9\textrm{ GeV}, & \qquad m_b = 4.79\textrm{ GeV}, \qquad m_\tau = 1.777\textrm{ GeV}\;.
	\nonumber
\end{align}

\begin{table}[t]\footnotesize
\begin{tabular}{|c|c|c|}
\hline
& Notation & Measurement \\
\hline
Z-pole & $\Gamma_Z$ & Total $Z$ width \\
& $\sigma_{\rm had}$ &	Hadronic cross section \\
& $R_f$($f=e,\mu,\tau,b,c$) & Ratios of decay rates \\
& $A_{FB}^{0,f}$($f=e,\mu,\tau,b,c,s$) & Forward-backward asymmetries \\
& $\bar{s}_l^2$ & Hadronic charge asymmetry \\
& $A_f$($f=e,\mu,\tau,b,c,s$) & Polarized asymmetries \\
\hline
Fermion pair & $\sigma_f$($f=q,e,\mu,\tau)$ & Total cross sections for $e^+e^-\rightarrow f\bar{f}$ \\
production at LEP2 & $A_{FB}^f$($f=\mu,\tau$) & Forward-backward asymmetries for $e^+e^-\rightarrow f\bar{f}$ \\
\hline
$W$ mass & $m_W$ & $W$ mass from LEP and Tevatron \\
and decay rate & $\Gamma_W$ & $W$ width from Tevatron \\
\hline
DIS & $Q_W(Cs)$	& Weak charge in Cs \\
and & $Q_W(Tl)$ & Weak charge in Tl \\
atomic parity violation	& $Q_W(e)$ & Weak charge of the electron \\
& $g_L^2,g_R^2$	& $\nu_\mu$-nucleon scattering from NuTeV \\
& $g_V^{\nu e},g_A^{\nu e}$ & $\nu$-$e$ scattering from CHARM II \\
\hline
\end{tabular}
\caption{Precision electroweak quantities.  Data taken from \cite{Beringer:1900zz,Alcaraz:2006mx}.\label{tab:observables}}
\end{table}

We now apply equations (\ref{dx1}) and (\ref{dx2}) to the electroweak observables listed in Table \ref{tab:observables}.  In order to obtain a bound on the coefficients $c_{WW}/\Lambda^2$ and $c_{BB}/\Lambda^2$, we use the $\chi^2$ statistic
\begin{equation}
	\chi^2 = \sum_{i,j} \chi^i \left(\sigma^{-1}\right)_{ij} \chi^j
\end{equation}
where $\chi^i = \left(X^i_{SM}-X^i_{exp}+\frac{c_{WW}}{\Lambda^2}X^i_{WW}
+\frac{c_{BB}}{\Lambda^2}X^i_{BB} +\frac{c_{BW}}{\Lambda^2}X^i_{BW}\right)$ and $\sigma_{ij}$ is the error matrix, related to the errors for each observable, $\sigma_i$, and the error correlation matrix, $\rho_{ij}$,\cite{Beringer:1900zz,Alcaraz:2006mx}
\begin{equation}
	\sigma_{ij} = \sigma_i \rho_{ij} \sigma_j\;.
\end{equation}

We calculate the bounds by first setting $c_{BW}$ to the value (as a function of $c_{WW}$ and $c_{BB}$) which minimizes $\chi^2$, $c_{BW}^{min}$.  Thus we make no assumptions about $c_{BW}$, but rather allow its value to float.  We then write this new $\chi^2$ in the following way
\begin{equation}
	\chi^2\big|_{c_{BW}=c_{BW}^{min}} = \chi^2_{min} +
	\frac{(c_i-\hat{c}_i)M_{ij}(c_j-\hat{c}_j)}{\Lambda^4} \qquad i,j \in \{WW, BB\}
\end{equation}
where $\chi^2_{min}$ is the value of $\chi^2$ minimized with respect to all coefficients, $\hat{c}_i$ is the best fit value of the coefficient $c_i$, and $M_{ij}$ is a symmetric matrix.  We arrive at bounds by solving the equation
\begin{equation}
	\frac{(c_i-\hat{c}_i)M_{ij}(c_j-\hat{c}_j)}{\Lambda^4} = 1
\end{equation}

We can diagonalize $M_{ij}$ to find two statistically independent combinations of our two operators and obtain a bound on those.  We find
\begin{equation}
	\begin{pmatrix}
		0.999 & 0.0385 \\
		-0.0385 & 0.999
	\end{pmatrix} \times \frac{1}{\Lambda^2}
	\begin{pmatrix}
		c_{WW} \\
		c_{BB}
	\end{pmatrix}
	= \begin{pmatrix}
	129.4 \pm 120.7 \text{ TeV}^{-2} \\
	-482.3 \pm 3160 \text{ TeV}^{-2}
	\end{pmatrix}
\end{equation}
The contributions of the two coefficients are essentially decoupled due to the fact that the net effect of $\mathcal{O}_{WW}$ is significantly larger than that of $\mathcal{O}_{BB}$.  While both coefficients are consistent with zero, the central value of $\mathcal{O}_{WW}$ differs from zero by just over one standard deviation.

If we instead compute bounds for each coefficient separately, setting the other coefficient to zero in each case, we obtain
\begin{align}
	\frac{c_{WW}}{\Lambda^2} & = 129.5 \pm 120.8 \text{ TeV}^{-2} \\
	\frac{c_{BB}}{\Lambda^2} & = 1456 \pm 2225 \text{ TeV}^{-2}
\end{align}
We again see that the net effect of $\mathcal{O}_{WW}$ is significantly larger than that of $\mathcal{O}_{BB}$.  

We can attempt to gain some understanding of the very different numerical impact of these two operators by considering the oblique parameters $\hat S,\hat T,\hat U,\hat V,\hat W,\hat X,\hat Y$ of Ref.~\cite{Barbieri:2004qk}.  These parameters are defined in terms of a Taylor expansion of the self energies about $q^2=0$, and are therefore useful if the values of $q^2$ probed by the experiments are within the radius of convergence of the expansion.  The nearest singularities of the self energies are branch points at $(M_W + m_h)^2$ and $(M_Z + m_h)^2$ from the first two diagrams in Fig.~1.  Since all precision electroweak data are at values of $q^2$ less than these values, the Taylor expansion should converge.  In a sense, the Higgs boson is playing the role of the ``heavy'' particle in the formalism, analogous to the role of the top quark in Ref.~\cite{Greiner:2011tt}.

Analytic expressions for the oblique parameters $\hat S,\hat T,\hat U,\hat V,\hat W,\hat X,\hat Y$ are given in Appendix~\ref{apdx:oblique}.  We find that the coefficient $c_{BB}/\Lambda^2$ contributes only to $\hat S$, $\hat X$, and $\hat Y$.  Because $\hat S$ contains a contribution at tree level from $c_{BW}/\Lambda^2$, it cannot be used to constrain $c_{BB}/\Lambda^2$; we conclude that $\hat X$ and $\hat Y$ must be the oblique parameters that determine the bound on $c_{BB}/\Lambda^2$.  Unfortunately, $\hat X$ and $\hat Y$ are not parametrically suppressed with respect to the other oblique parameters, so there is no obvious reason why the bound on $c_{BB}/\Lambda^2$ is so much weaker than the bound on $c_{WW}/\Lambda^2$.  

This marks the first time bounds on these operators have been obtained from precision electroweak data using the full power of the effective field theory framework.  The bounds obtained are unambiguous and independent of any assumptions about the coefficient of the operator $\mathcal{O}_{BW}$, which was allowed to float.  An analogous result for top-quark loops is given in Refs.~\cite{Greiner:2011tt, Zhang:2012cd}.

Let us compare this calculation with the previous calculations of Refs.~\cite{De Rujula:1991se,Hagiwara:1993ck,Alam:1997nk}.  As mentioned above, these papers did not appreciate that unambiguous bounds could be obtained on the operators from a one-loop analysis of precision electroweak data.  In Ref.~\cite{Hagiwara:1993ck,Alam:1997nk}, only the divergent portion of the one-loop diagrams, and a subset of the finite parts (those that are proportional to $\ln m_h$ or $m_h^2$), were calculated.  A bound on the coefficients of $\mathcal{O}_{WW}$ and $\mathcal{O}_{BB}$ was obtained by assuming that the coefficient of the operator $\mathcal{O}_{BW}$ vanishes.  This is an unjustified assumption, because the coefficient of the operator $\mathcal{O}_{BW}$ is renormalized by $\mathcal{O}_{WW}$ and $\mathcal{O}_{BB}$.  Explicitly, in the $\overline{\rm MS}$ scheme,
\begin{equation}
	c_{BW}(\mu) = c_{BW}^{0} - \frac{g^2(\mu)}{16\pi^2}\left(c_{WW} + \frac{s^2}{c^2}c_{BB}\right)
	\left(\frac{1}{\epsilon}-\gamma+\ln 4\pi\right)
\end{equation}
where $c_{BW}(\mu)$ is the renormalized coefficient and $c_{BW}^{0}$ is the bare coefficient.  There is no reason why the renormalized coefficient should vanish.  Making this unjustified assumption allows one to extract much stronger bounds on $c_{WW}$ and $c_{BB}$, but these stronger bounds are specious.  In contrast, our bounds are obtained by letting $c_{BW}(\mu)$ float.  The resulting bounds on $c_{WW}$ and $c_{BB}$ are unambiguous and do not depend on the renormalization scale $\mu$.

The bounds we obtained on the coefficients of the operators $\mathcal{O}_{WW}$ and $\mathcal{O}_{BB}$ at one loop from precision electroweak data are so weak that they are easily eclipsed by bounds obtained by tree-level processes involving the Higgs boson \cite{GonzalezGarcia:1998wn,GonzalezGarcia:1999fq,Corbett:2012dm,Corbett:2012ja,Masso:2012eq,Chang:2013cia}.
Thus we have shown that, when done properly, the bounds on operators obtained from a one-loop analysis of precision electroweak data cannot compete with bounds obtained from tree-level processes, as one would naturally suspect.

\appendix
\section{Self Energies}
\label{apdx:corrections}

The expressions below are given in terms of scalar integral functions, $A_0$ and $B_0$.  Expressions for these functions are given in Appendix D of Ref.~\cite{Passarino:1978jh}.

\vspace{24pt}
\noindent $\mathcal{O}_{BB}$:
\nopagebreak
\vspace{14pt}
\nopagebreak
\begin{align*}
	\Pi_{WW} \quad & = \quad \frac{c_{BB}}{\Lambda^2} \frac{1}{16\pi^2} \frac{g^2 m_Z^4 \sin^4\theta_W}{m_h^2}\left(2m_Z^2-3 A_0\left(m_Z^2\right)\right) \\
	\Pi_{ZZ} \quad & = \quad \frac{c_{BB}}{\Lambda^2} \frac{1}{16\pi^2} \frac{g^2 \sin^4\theta_W}{m_h^2 \cos^2\theta_W}\left[2 m_h^2 m_Z^2\left(q^2-m_h^2 + m_Z^2\right)B_0\left(q^2,m_h^2,m_Z^2\right)\right. \\
	& \qquad - m_Z^2 \left(3 q^2  + 2 m_h^2 + 3m_Z^2\right)A_0\left(m_Z^2\right) + m_h^2 \left(2m_Z^2 - q^2\right) A_0\left(m_h^2\right) \\
	& \qquad \left. - 6 m_W^2 q^2 A_0\left(m_W^2\right) + 4m_W^4 q^2 + 2 m_Z^4 q^2 + 2m_Z^6\right] \\
	\Pi_{\gamma\gamma} \quad & = \quad -\frac{c_{BB}}{\Lambda^2} \frac{1}{16\pi^2} \frac{g^2 q^2 \sin^2\theta_W}{m_h^2}\left[m_h^2 A_0\left(m_h^2\right) + 3m_Z^2 A_0\left(m_Z^2\right) \right. \\
	& \qquad \left. + 6m_W^2 A_0\left(m_W^2\right) -4m_W^4 - 2m_Z^4\right] \\
	\Pi_{\gamma Z} \quad & = \quad \frac{c_{BB}}{\Lambda^2} \frac{1}{16\pi^2} \frac{g^2 \sin^3\theta_W}{m_h^2 \cos\theta_W} \left[m_h^2 m_Z^2\left(m_h^2 - m_Z^2 - q^2\right)B_0\left(q^2,m_h^2,m_Z^2\right)\right. \\
	& \qquad + m_Z^2\left(m_h^2 + 3q^2\right) A_0\left(m_Z^2\right) - m_h^2\left(m_Z^2-q^2\right) A_0\left(m_h^2\right) \\
	& \qquad \left. + 6 m_W^2 q^2 A_0\left(m_W^2\right) - 4m_W^4 q^2 - 2m_Z^4 q^2\right]
\end{align*}

\vspace{20pt}
\noindent $\mathcal{O}_{WW}$:
\nopagebreak
\vspace{14pt}
\nopagebreak
\begin{align*}
	\Pi_{WW} \quad & = \quad \frac{c_{WW}}{\Lambda^2} \frac{1}{16\pi^2} \frac{g^2}{m_h^2} \left[2m_h^2 m_W^2 \left(q^2-m_h^2+m_W^2\right) B_0\left(q^2,m_h^2,m_W^2\right)\right. \\
	& \qquad - 2m_W^2\left(3q^2 + m_h^2 + 3m_W^2\right) A_0\left(m_W^2\right) + m_h^2\left(2m_W^2 - q^2\right)A_0\left(m_h^2\right) \\
	& \qquad \left. - 3\left(m_W^4 + m_Z^2 q^2\right) A_0\left(m_Z^2\right) + 4m_W^4 q^2 + 2m_Z^4 q^2 + 4m_W^6 + 2m_W^4 m_Z^2\right] \\
	\Pi_{ZZ} \quad & = \quad \frac{c_{WW}}{\Lambda^2} \frac{1}{16\pi^2} \frac{g^2 \cos^2\theta_W}{m_h^2} \left[2m_h^2 m_Z^2\left(q^2 - m_h^2 + m_Z^2\right) B_0\left(q^2,m_h^2,m_Z^2\right) \right. \\
	& \qquad - m_Z^2\left(3q^2 + 2m_h^2 + 3m_Z^2\right) A_0\left(m_Z^2\right) + m_h^2 \left(2m_Z^2 - q^2\right) A_0\left(m_h^2\right) \\
	& \qquad \left. - 6\left(m_Z^4 + m_W^2 q^2\right) A_0\left(m_W^2\right) + 4 m_W^4 q^2 + 2 m_Z^4 q^2 + 4m_W^2 m_Z^4 + 2 m_Z^6\right] \\
	\Pi_{\gamma\gamma} \quad & = \quad -\frac{c_{WW}}{\Lambda^2} \frac{1}{16\pi^2} \frac{g^2 q^2 \sin^2\theta_W}{m_h^2} \left[m_h^2 A_0\left(m_h^2\right) + 3 m_Z^2 A_0\left(m_Z^2\right) \right. \\
	& \qquad \left. + 6 m_W^2 A_0\left(m_W^2\right) - 4m_W^4 - 2m_Z^4\right] \\
	\Pi_{\gamma Z} \quad & = \quad -\frac{c_{WW}}{\Lambda^2} \frac{1}{16\pi^2} \frac{g^2 \sin\theta_W \cos\theta_W}{m_h^2} \left[m_h^2 m_Z^2\left(m_h^2 - m_Z^2 - q^2\right) B_0\left(q^2,m_h^2,m_Z^2\right) \right. \\
	& \qquad + m_Z^2\left(m_h^2 + 3q^2\right)A_0\left(m_Z^2\right) - m_h^2 \left(m_Z^2 - q^2\right) A_0\left(m_h^2\right) \\
	& \qquad \left. + 6 m_W^2 q^2 A_0\left(m_W^2\right) - 4 m_W^4 q^2 - 2 m_Z^4 q^2\right]
\end{align*}

\section{Oblique Parameters}
\label{apdx:oblique}

\begin{align}
	\hat{S} & = -\frac{c_{BW}(\mu)}{\Lambda^2} m_W^2 + \frac{g^2 m_W^2}{32\pi^2(m_h^2 -
	 m_Z^2)^2}\left(\frac{c_{WW}}{\Lambda^2} + \frac{s^2}{c^2}\frac{c_{BB}}{\Lambda^2}\right) \\
	& \qquad \times \left(-2m_h^2(m_h^2-2m_Z^2) \ln\left(\frac{m_h^2}{\mu^2}\right) 
	- 2m_Z^4 \ln\left(\frac{m_Z^2}{\mu^2}\right)
	 + (m_h^2 - 3m_Z^2)(m_h^2-m_Z^2)\right) \nonumber \\
	\hat{T} & = 0 \\
	\hat{U} & = \frac{g^2 m_W^2 m_Z^2}{8\pi^2(m_h^2-m_Z^2)^2(m_h^2-m_W^2)^2} \frac{c_{WW}}{\Lambda^2}
	\left(-m_h^2 m_Z^2 s^2\left(m_h^2(s^2-2)+2m_W^2\right)\ln\left(\frac{m_h^2}{\mu^2}\right) \right. \\
	& \qquad \left. - m_Z^2(m_h^2-m_W^2)^2 \ln\left(\frac{m_Z^2}{\mu^2}\right) + m_W^2 c^2 (m_h^2-m_Z^2)^2
	\ln\left(\frac{m_W^2}{\mu^2}\right) - m_h^2 s^2(m_h^2-m_Z^2)(m_h^2-m_W^2)\right) \nonumber \\
	\hat{V} & = \frac{g^2 m_W^4 m_Z^2}{24\pi^2(m_h^2-m_Z^2)^4(m_h^2-m_W^2)^4}\frac{c_{WW}}{\Lambda^2}
	\left(-6m_h^2 m_W^2 c^2(m_h^2-m_Z^2)^4 \ln\left(\frac{m_W^2}{m_h^2}\right) \right. \\
	& \qquad \left. + 6m_h^2 m_Z^2 
	(m_h^2-m_W^2)^4 \ln\left(\frac{m_Z^2}{m_h^2}\right) \right. \nonumber \\
	& \qquad \left. + s^2(m_h^2-m_Z^2)(m_h^2-m_W^2) 
	(2m_h^8+5m_h^6(m_Z^2+m_W^2)-m_h^4(m_W^4+22m_W^2 m_Z^2 + m_Z^4) 
	\nonumber \right. \\
	& \qquad\qquad \left. +5m_h^2 m_W^2 m_Z^2(m_W^2+m_Z^2)
	+ 2m_W^4 m_Z^4)\vphantom{\frac12}\right) \nonumber \\
	\hat{W} & = \frac{g^2 m_W^4}{24\pi^2(m_h^2-m_Z^2)^4}\frac{c_{WW}}{\Lambda^2}
	\left(6m_h^2 m_Z^4 \ln\left(\frac{m_Z^2}{m_h^2}\right) - (m_h^2 - m_Z^2)
	(m_h^4 - 5m_h^2 m_Z^2 - 2m_Z^4)\right) \\
	\hat{X} & = \frac{g^2 m_W^4}{48\pi^2(m_h^2-m_Z^2)^4} 
	\frac{s}{c}\left(\frac{c_{WW}}{\Lambda^2} + \frac{s^2}{c^2}\frac{c_{BB}}{\Lambda^2}\right) \\
	& \qquad \times \left(-6m_h^2 m_Z^4 \ln\left(\frac{m_Z^2}{m_h^2}\right) 
	+ (m_h^2-m_Z^2)(m_h^4-5m_h^2 m_Z^2 - 2m_Z^4)\right) \nonumber \\
	\hat{Y} & = \frac{g^2 m_Z^2 s^2}{24\pi^2 (m_h^2 - m_Z^2)^4} \frac{c_{BB}}{\Lambda^2}
	\left(6m_h^2 m_Z^6 s^2 \ln\left(\frac{m_Z^2}{m_h^2}\right) - m_Z^2 s^2(m_h^2 - m_Z^2)
	(m_h^4 - 5m_h^2 m_Z^2 - 2m_Z^4)\right)
\end{align}

Large $m_h$ limit:
\begin{align}
	\hat{S} & = -\frac{c_{BW}(\mu)}{\Lambda^2} m_W^2 
	+ \frac{g^2 m_W^2}{32\pi^2}\left(\frac{c_{WW}}{\Lambda^2} 
	+ \frac{s^2}{c^2}\frac{c_{BB}}{\Lambda^2}\right)
	\left(1-2\ln\left(\frac{m_h^2}{\mu^2}\right)\right) \\
	\hat{U} & = -\frac{g^2 s^2 m_W^2 m_Z^2}{8\pi^2 m_h^2} \frac{c_{WW}}{\Lambda^2} \\
	\hat{V} & = \frac{g^2 s^2 m_W^4 m_Z^2}{12\pi^2 m_h^4} \frac{c_{WW}}{\Lambda^2} \\
	\hat{W} & = -\frac{g^2 m_W^4}{24\pi^2 m_h^2} \frac{c_{WW}}{\Lambda^2} \\
	\hat{X} & = \frac{g^2 m_W^4}{48\pi^2 m_h^2} \frac{s}{c} \left(\frac{c_{WW}}{\Lambda^2} +
	\frac{s^2}{c^2} \frac{c_{BB}}{\Lambda^2}\right) \\
	\hat{Y} & = -\frac{g^2 s^4 m_Z^4}{24\pi^2 m_h^2} \frac{c_{BB}}{\Lambda^2}
\end{align}

\end{document}